\documentclass[manuscript]{aastex}

\usepackage{amsmath}

\def\pmb#1{\mbox{\boldmath$#1$}}
\def\pmbmt#1{\pmb{\sf #1}}

\def\pmb#1{\mbox{\boldmath$#1$}}

\def\gtsim {\gtrsim}

\def\be{\begin{equation}}
\def\ee{\end{equation}}
\def\be{\begin{eqnarray}}
\def\ee{\end{eqnarray}}

\def\pmbmt#1{\pmb{\sf #1}}
\def\rmi{{\rm i}}

\shorttitle{Rotating Hot Jupiters}
\shortauthors{U.Lee}

\begin{document}

\title{Overstable Convective Modes of Rotating Hot Jupiters}

\author{Umin \textsc{Lee}}
\email{lee@astr.tohoku.ac.jp}
\affil{Astronomical Institute, Tohoku University, Sendai, Miyagi 980-8578, Japan}

\date{Accepted XXX. Received YYY; in original form ZZZ}


\begin{abstract}
We calculate overstable convective modes of uniformly rotating hot Jupiters, which
have a convective core and a thin radiative envelope.
Convective modes in rotating planets have complex frequency $\omega$ and 
are stabilized by rapid rotation so that their growth rates $\propto\omega_{\rm I}={\rm Im}(\omega)$ are much smaller than those for the non-rotating planets.
The stabilized convective modes excite low frequency gravity waves in the radiative envelope by frequency resonance between them.
We find that the convective modes that excite envelope gravity waves remain unstable even in the presence of non-adiabatic dissipations in the envelope.
We calculate the heating rates due to non-adiabatic dissipations of the oscillation energy of the unstable convective modes and find that the magnitudes of the heating rates cannot be large enough to inflate hot Jupiters sufficiently so long as the oscillation amplitudes remain in the linear regime.

\end{abstract}


\section{Introduction}

Calculating adiabatic convective modes of uniformly rotating massive main sequence stars, Lee \& Saio (1986) found 
that the convective modes of rotating stars become overstable and are stabilized by rapid rotation
so that their growth rates are much smaller than those of the non-rotating stars.
They also found that the stabilized convective modes excite high radial order $g$-modes in the envelope of the stars when the convective modes are in frequency resonance with envelope $g$-modes.
Carrying out non-adiabatic calculations, Lee \& Saio (1987b) also showed that the unstable convective modes that excite envelope $g$-modes remain unstable even in the presence of non-adiabatic dissipations in the envelope.
Using the traditional approximation (e.g., Lee \& Saio 1987a, 1997), it was shown that the rotationally stabilized convective modes have a branch of negative oscillation energy (Lee \& Saio 1990b). 
The excitation of envelope $g$-modes by the convective modes is caused by
linear mode coupling between the convective modes having negative energy of oscillation and
$g$-modes, which have positive oscillation energy (Lee \& Saio 1989).

It has been suggested that hot Jupiters orbiting close to the host stars
have larger radii compared with those of Jovian planets of similar masses and ages, orbiting far from their host stars.
The strong irradiation by the host star, however, is not necessarily effective to inflate the planets to the radii observationally estimated (e.g., Baraffe et al. 2003).
To explain the larger radii of the hot Jupiters, tidal heating in the interior of the planets has been suggested.
However, it is well known that gravitational tides are likely to cause synchronization between the 
orbital motion and spin of the planet in timescales much shorter than the ages of the planets
(e.g., Bodenheimer et al 2001).
Some authors have proposed thermal tides as a mechanism that keeps the spin asynchronous with the orbital motion of the planets
so that tidal heating in the interior can be operative
(e.g., Arras \& Socrates 2010; Auclair-Desrotour et al 2017; Auclair-Desrotour \& Leconte 2018).

We compute overstable convective modes of rotating hot Jupiters.
Overstable convective modes in rotating Jovian planets could produce various low frequency phenomena in the planets.
Lee \& Saio (1990a) proposed that $g$-modes excited in the Jovian atmosphere by 
overstable convective modes could be a cause of low frequency phenomena observed in Jupiter (e.g., Deeming et al 1989; Magalh\~aes et al 1989).
In this paper, as a heating mechanism, 
we estimate the amount of non-adiabatic dissipation of the oscillation energy of
unstable convective modes of hot Jupiters.
Here, we consider unstable convective modes in the core which excite gravity waves in the envelope,
expecting that
this excitation of $g$-modes could enhance thermal dissipation in the envelope.
To estimate the dissipation rates of the oscillation energy, we need non-adiabatic
treatments of the oscillations of rotating planets.
For this purpose we employ an approximate treatment of non-adiabatic oscillations, 
following Auclair-Desrotour \& Leconte (2018).
Method of solution for the non-adiabatic calculations is given in \S 2 and the numerical results
and conclusions are presented in \S 3 and \S 4, respectively.
We give a brief account of the traditional approximation applied to the stabilized convective modes in the Appendix.

\section{Method of Solutions}

\subsection{Equilibrium Model}

Following Arras \& Socrates (2010) and Auclair-Desrotour \& Leconte (2018), 
we employ for our modal analyses of hot Jupiters simple static models, 
consisting of a thin radiative envelope and a convective core.
In this paper, we ignore the deformation of rotating planets and hence the planet models are 
assumed to be spherical symmetric.
Such Jovian models are computed by integrating the hydrostatic equations
\be
{dp\over dr}=-\rho{GM_r\over r^2},
\ee
\be
{dM_r\over dr}=4\pi r^2\rho,
\ee
with the equation of state given by
\be
\rho(p)=e^{-p/p_b}{p\over a^2}+\left(1-e^{-p/p_b}\right)\sqrt{p\over K_c},
\ee
where $p$ is the pressure, $\rho$ is the mass density, $G$ is the gravitational constant, 
$p_b$ is the pressure at the base of the radiative envelope,
and
\be
K_c=GR_J^2, \quad a^2=\sqrt{p_bK_c},
\ee
and $a$ is the isothermal sound velocity, and $R_J$ is the radius of Jupiter.
In the convective core where $p\gg p_b$, we obtain
\be
p\approx K_c\rho^{\Gamma} \quad {\rm with} \quad \Gamma=2,
\ee
corresponding to a polytrope of the index $n=1$, and hence for $\Gamma_1\equiv(\partial\ln p/\partial\ln \rho)_S\approx 2$ the square of the Brunt-V\"ais\"al\"a frequency 
\be
N^2\equiv -gA\approx 0,
\ee
where 
\be
g=-{1\over\rho}{dp\over dr}={GM_r\over r^2},
\quad A\equiv{d\ln\rho\over dr}-{1\over\Gamma_1}{d\ln p\over dr}. 
\ee
On the other hand, in the radiative envelope where $p\ll p_b$, we have
\be
p\approx a^2\rho,
\ee
and
\be
N^2\approx 
(\Gamma_1-1){g^2\over c^2}, \quad c^2=\Gamma_1{p\over\rho},
\ee
and $c$ denotes the adiabatic sound speed.
Since $p/\rho\approx a^2 \propto T$ for an ideal gas, 
the temperature $T$ is constant for a constant $a$, suggesting an isothermal atmosphere.

In this paper, following Auclair-Desrotour \& Leconte (2018),
for the jovian planet model of mass $M=0.7M_J$ with $M_J$ being the Jovian mass, 
we use $p_b=100{\rm bar}=10^8{\rm dyn/cm^2}$ and set the outer boundary $R_e$ at
$p=0.01{\rm dyn/cm^2}$ and define the planet's radius $R=R_e/1.01=9.31\times10^9{\rm cm}$.
We also assume $\Gamma_1=2$ in the interior.
Figure 1 is the propagation diagram where $N^2$ and $L_l^2\equiv l(l+1)c^2/r^2$ with $l=2$ are plotted as a function of $r/R$ where $L_l$ is 
the Lamb frequency.

\begin{figure}
\resizebox{0.45\columnwidth}{!}{
\includegraphics{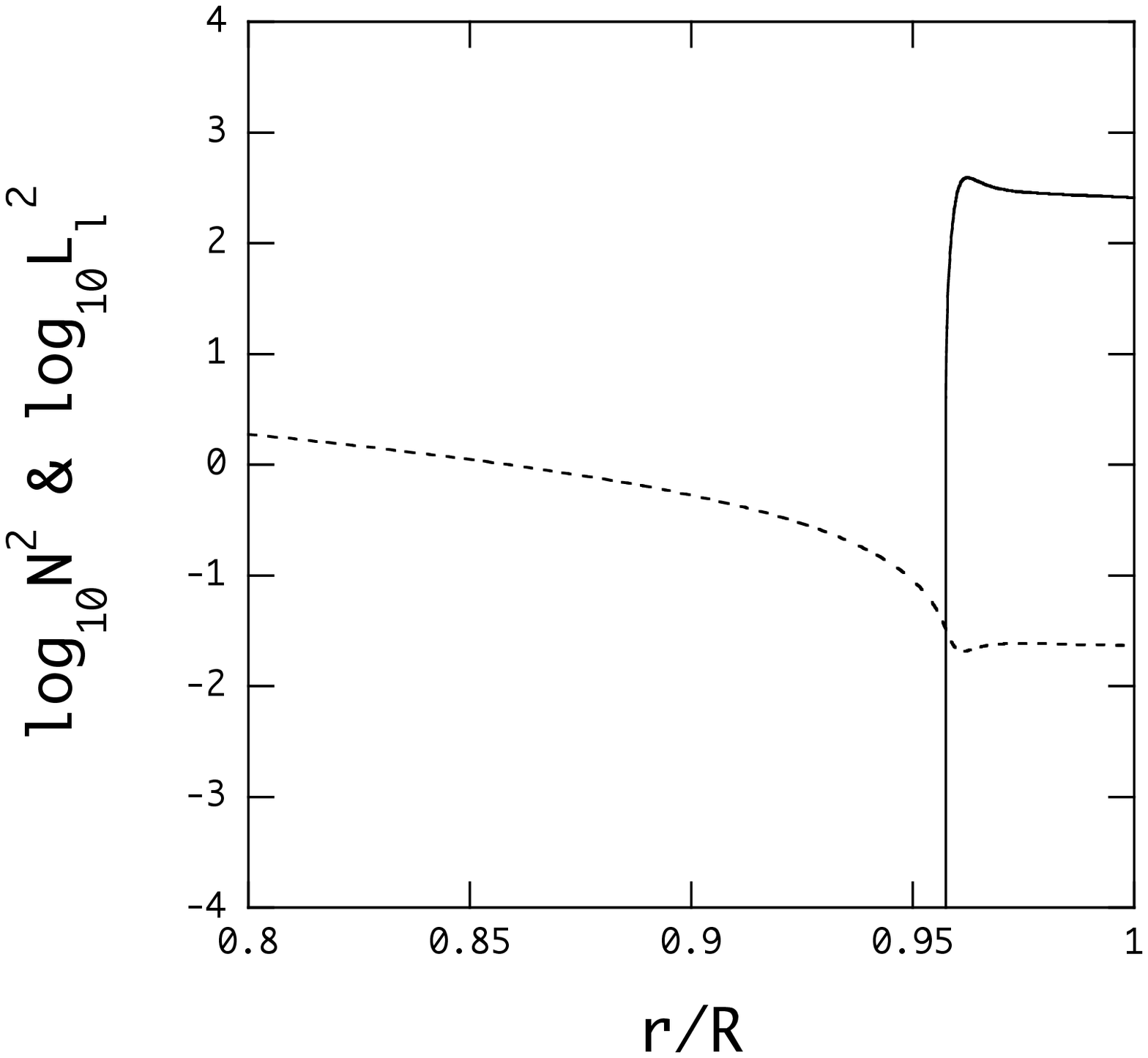}}
\resizebox{0.45\columnwidth}{!}{
\includegraphics{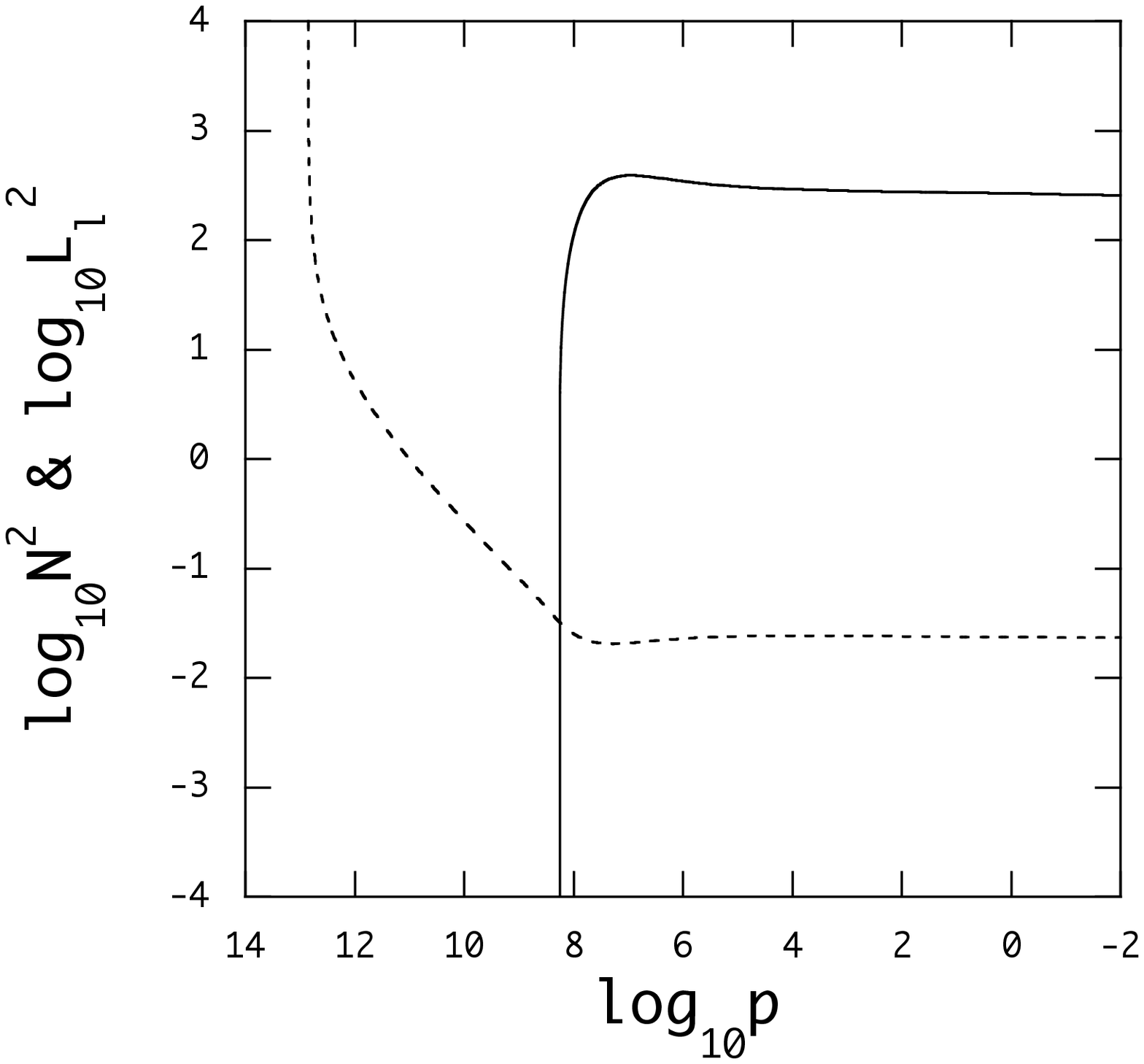}}
\caption{Squared characteristic frequencies $N^2$ (solid lines) and $L_l^2$ (dashed lines) plotted as a function of $r/R$ (left panel)
and of $\log p$ (right panel), where $N$ and $L_l$ are respectively the Brunt-V\"ais\"al\"a frequency and Lamb frequency. Note that
the squared frequencies are normalized by $GM/R^3$ where $M$ and $R$ are the mass and radius of the planet,
$G$ is the gravitational constant and
$l=2$ is used for $L_l$.}
\end{figure}

\subsection{Perturbed Basic Equations}

The perturbed basic equations for a uniformly rotating planet may be given by
\be
-\rho\omega^2\pmb{\xi}+2\rmi\rho\omega\pmb{\Omega}\times\pmb{\xi}-{\rho'\over\rho}\nabla p+\nabla p'=0,
\ee
\be
\rho'+\nabla\cdot\left(\rho\pmb{\xi}\right)=0,
\ee
\be
\rmi\omega\rho T\delta s=-\nabla\cdot\pmb{F}',
\label{eq:dsdt}
\ee
\be
{\delta s\over c_p}={1\over\alpha_T}\left({1\over\Gamma_1}{\delta p\over p}-{\delta\rho\over\rho}\right),
\label{eq:deltas}
\ee
where
$\rho$, $p$, $s$, $T$, and $c_p$ are respectively the mass density, pressure, specific entropy,
temperature, and specific heat at constant pressure, $\pmb{F}$ is the energy flux vector, $\pmb{\xi}$ is the displacement vector, 
$
\alpha_T=-\left({\partial\ln\rho/\partial\ln T}\right)_P,
$
and $(')$ and $\delta$ indicate resepctively the Eulerian and Lagrangian perturbation of quantities.
Note that we have applied the Cowling approximation neglecting the Euler perturbation of the gravitational potential due to self-gravity, and that
we have ignored the effects of rotational deformation.
The vector $\pmb{\Omega}$ denotes the angular velocity vector of rotation and is assumed to be
constant for uniformly rotating planets.
Here, we have assumed that the perturbations are proportional to the factor ${\rm e}^{\rmi\omega t+\rmi m\phi}$, 
where $\omega=\sigma+m\Omega$ with $\Omega=|\pmb{\Omega}|$ is the oscillation frequency in
the co-rotating frame of the planet and $\sigma$ is the frequency observed in the inertial frame, and
$m$ denotes the azimuthal wavenumber around the axis of rotation.

Since evolutionary models of irradiated Jovian planets computed with appropriate equations of state
and opacities are not available to us,
we employ an approximate treatment to estimate the non-adiabatic effects $\delta s$,
following Auclair-Desrotour \& Leconte (2018).
We write the term on the right-hand-side of equation (\ref{eq:dsdt}) as
\be
\nabla\cdot\pmb{F}'
\sim { \omega_D\rho c_p T}{T'\over T}
={\omega_D \rho c_p T}\left({\delta T\over T}+\nabla V{\xi_r\over r}\right)=
{\omega_D\rho c_p T}\left({\delta s\over c_p}+\nabla_{ad}{\delta p\over p}+\nabla V{\xi_r\over r}\right),
\ee
where we have used
$\delta s/c_p=\delta T/T-\nabla_{ad}\delta p/p$ with $\nabla_{ad}=(\partial\ln T/\partial\ln p)_S$
and $\nabla={d\ln T/ d\ln p}$, 
and  
$
\omega_D\equiv {2\pi/ \tau_0}.
$
Here, $\tau_0$ is the thermal timescale given by
\be
\tau_0=\tau_*\times{1\over 2}\left[\left({p\over p_*}\right)^{1/2}+\left({p\over p_*}\right)^2\right],
\ee
where $p_*(\ll p_b)$ is the pressure 
at the base of the heated layer by the irradiation of the host stars
and is set equal to $p_*=10^6{\rm dyne/cm^2}$ (e.g., Auclair-Desrotour \& Leconte 2018), and $\tau_*$ is a parameter to specify the efficiency
of the radiative cooling
and $\tau_*=1{\rm day}$ is assumed (Auclair-Desrotour \& Leconte 2018; see also Iro et al 2005).
From equation (\ref{eq:dsdt}) 
we obtain
\be
{\delta s\over c_p}=-{\omega_D\over\rmi\omega+\omega_D}\left(\nabla_{ad}{\delta p\over p}+\nabla V{\xi_r\over r}\right).
\label{eq:deltas2}
\ee
When we take the limit of short thermal timescales $\tau_0\rightarrow 0$ (i.e.,
$\omega_D\rightarrow \infty$),  
we obtain isothermal perturbations $ T'/T=0$.
On the other hand, in the limit of long thermal timescale $\tau_0\rightarrow \infty$
so that $\omega_D\rightarrow 0$, we have $\delta s\rightarrow 0$, which corresponds to adiabatic perturbations, expected in the deep interior
of the planet.

\subsection{Oscillation Equations}

To describe the perturbations of rotating planets, we employ a series expansion in terms of spherical harmonic functions
$Y_l^m(\theta,\phi)$ for a given $m$ with different $l$s, assuming the planet is axisymmetric.
The pressure perturbation, for example, is given as
\be
p'(r,\theta,\phi,t)=\sum_lp'_l(r)Y_l^m(\theta,\phi)e^{\rmi\omega t}, 
\ee
and the displacement vector $\pmb{\xi}$ is
\be
\xi_r(r,\theta,\phi,t)=r\sum_{l}S_l(r)Y_l^m(\theta,\phi)e^{\rmi\omega t},
\ee
\be
\xi_\theta(r,\theta,\phi,t)=r\sum_{l,l'}\left[H_l(r){\partial\over\partial\theta}Y_l^m(\theta,\phi)+T_{l'}{1\over\sin\theta}{\partial\over\partial\phi}Y_{l'}^m(\theta,\phi)\right]e^{\rmi\omega t},
\ee
\be
\xi_\phi(r,\theta,\phi,t)=r\sum_{l,l'}\left[H_l(r){1\over\sin\theta}{\partial\over\partial\phi}Y_l^m(\theta,\phi)-T_{l'}{\partial\over\partial\theta}Y_{l'}^m(\theta,\phi)\right]e^{\rmi\omega t}.
\ee
where $l=|m|+2(j-1)$ and $l'=l+1$ for even modes and $l=|m|+2j-1$ and $l'=l-1$ for odd modes and
$j=1,~2,~\cdots,~j_{\rm max}$ (see, e.g., Lee \& Saio 1986).

Substituting the expansions into the perturbed basic equations we obtain a set of ordinary differential equations
for the expansion coefficients such as $S_l(r)$ and $p'_l(r)$.
Using the dependent variables defined as
\be
\pmb{y}_1=(S_l), \quad \pmb{y}_2=\left({p'_l\over\rho g r}\right), \quad \pmb{y}_6=\left({\delta s_l\over c_p}\right), \quad \pmb{h}=(H_l), \quad \pmb{t}=(T_{l'}),
\ee
we write the set of differential equations as
\be
r{d\pmb{y}_1\over dr}=\left({V\over\Gamma_1}-3\right)\pmb{y}_1-{V\over\Gamma_1}\pmb{y}_2+\pmbmt{\Lambda}_0\pmb{h}+\alpha_T\pmb{y}_6,
\label{eq:y1}
\ee
\be
r{d\pmb{y}_2\over dr}=(c_1\bar\omega^2+rA)\pmb{y}_1+(1-U-rA)\pmb{y}_2-2c_1\bar\omega\bar\Omega\left(m\pmb{h}+\pmbmt{C}_0\rmi\pmb{t}\right)
+\alpha_T\pmb{y}_6,
\label{eq:y2}
\ee
\be
-\pmbmt{M}_0\pmb{h}+\pmbmt{L}_1\rmi\pmb{t}=-\nu\pmbmt{K}\pmb{y}_1,
\label{eq:auxiliary1}
\ee
\be
\pmbmt{L}_0\pmb{h}-\pmbmt{M}_1\rmi\pmb{t}={1\over c_1\bar\omega^2}\pmb{y}_2+m\nu\pmbmt{\Lambda}_0^{-1}\pmb{y}_2,
\label{eq:auxiliary2}
\ee
\be
\pmb{y}_6=-{\omega_D\over\rmi\omega+\omega_D} V\left[\nabla_{ad}\pmb{y}_2-\left(\nabla_{ad}-\nabla\right)\pmb{y}_1 \right],
\label{eq:y6}
\ee
where 
\be
\nu={2\Omega\over\omega}, \quad \bar\omega={\omega\over\sigma_0}, \quad \bar\Omega={\Omega\over\sigma_0}, \quad \sigma_0=\sqrt{GM\over R^3},
\ee
\be
V=-{d\ln p\over d\ln r}, \quad U={d\ln M_r\over d\ln r}, \quad c_1={(r/R)^3\over M_r/M}, 
\ee
and $M$ and $R$ denote respectively the mass and radius of the planet, and
the temperature $T$ and the specific heat $c_p$ are assumed to be those for an ideal gas, that is,
\be
T={\mu\over{\cal R}}{p\over\rho}, \quad c_p={5\over 2}{{\cal R}\over\mu},
\ee
and $\cal R$ is the gas constant, and $\mu$ is the mean molecular weight, for which we use $\mu=1.3$, 
and the matrices $\pmbmt{\Lambda}_0$, $\pmbmt{C}_0$, $\pmbmt{L}_0$, $\pmbmt{L}_1$, $\pmbmt{M}_0$, $\pmbmt{M}_1$, $\pmbmt{K}$
are defined in Lee \& Saio (1990).

Eliminating the variables $\pmb{h}$, $\pmb{t}$, and $\pmb{y}_6$ between equations from (\ref{eq:y1}) to (\ref{eq:y6}), we obtain the oscillation equations given by
\be
r{d\pmb{y}_1\over dr}
=\left[\left({V\over\Gamma_1}-3+{\alpha_T (\nabla_{ad}-\nabla) V\over\rmi\omega/\omega_D+1}\right)\pmbmt{1}+\nu\pmbmt{W}\pmbmt{O}\right]\pmb{y}_1
+\left[{\pmbmt{W}\over c_1\bar\omega^2}-\left({V\over\Gamma_1}+{\alpha_T\nabla_{ad}V\over\rmi\omega/\omega_D+1}\right)\pmbmt{1}\right]\pmb{y}_2,
\label{eq:dy1dr}
\ee
\be
r{d\pmb{y}_2\over dr}
&=&\left[\left(c_1\bar\omega^2+rA+{\alpha_T(\nabla_{ad}-\nabla) V\over\rmi\omega/\omega_D+1}\right)\pmbmt{1}-4c_1\bar\Omega^2\pmbmt{G}\right]\pmb{y}_1\nonumber\\
&&+\left[\left(1-U-rA-{\alpha_T\nabla_{ad} V\over\rmi\omega/\omega_D+1}\right)\pmbmt{1}-\nu\pmbmt{O}^T\pmbmt{W}\right]\pmb{y}_2,
\label{eq:dy2dr}
\ee
where $\pmbmt{1}$ is the unit matrix, 
and the matrices $\pmbmt{W}$, $\pmbmt{O}$, and $\pmbmt{G}$ are
defined by
\be
\pmbmt{W}=\pmbmt{\Lambda}_0(\pmbmt{L}_0-\pmbmt{M}_1\pmbmt{L}_1^{-1}\pmbmt{M}_0)^{-1}, \quad 
\pmbmt{O}=m\pmbmt{\Lambda}_0^{-1}-\pmbmt{M}_1\pmbmt{L}_1^{-1}\pmbmt{K}, \quad 
\pmbmt{G}=\pmbmt{O}^T\pmbmt{W}\pmbmt{O}-\pmbmt{C}_0\pmbmt{L}_1^{-1}\pmbmt{K}.
\label{eq:wog}
\ee 
Note that by simply setting $\omega_D=0$, we restore the oscillation equations for adiabatic modes.
Because we assume the perturbations are proportional to ${\rm e}^{\rmi\omega t}$, modes with negative
$\omega_{\rm I}={\rm Im}(\omega)$ are unstable modes.

The boundary condition at the centre is the regularity condition of the functions $r\pmb{y}_1$
and $r\pmb{y}_2$.
We use the surface boundary conditions given by $\delta p=0$ at the surface.

\section{Numerical Results}

\begin{figure}
\resizebox{0.5\columnwidth}{!}{
\includegraphics{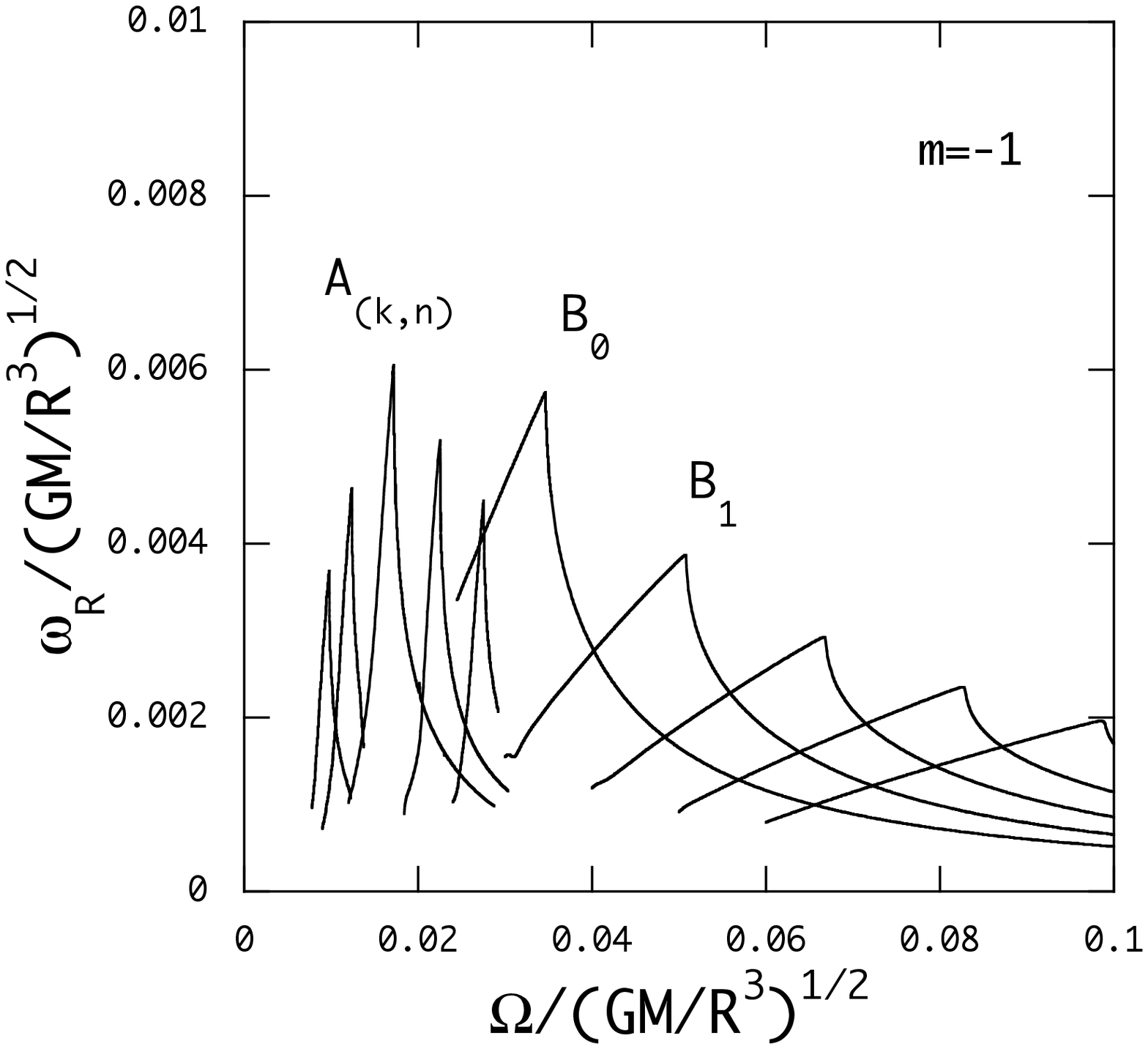}}
\resizebox{0.5\columnwidth}{!}{
\includegraphics{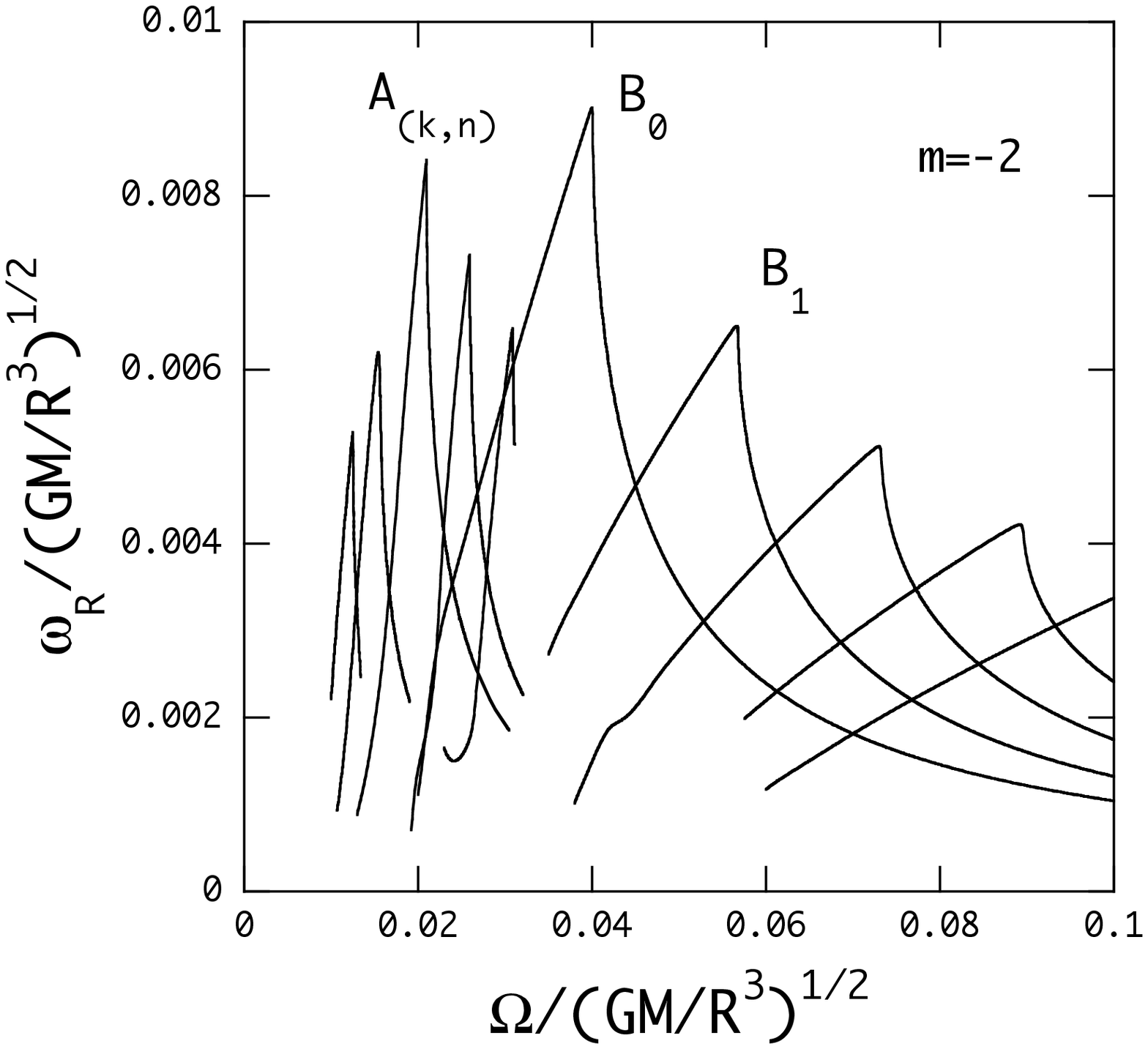}}
\resizebox{0.5\columnwidth}{!}{
\includegraphics{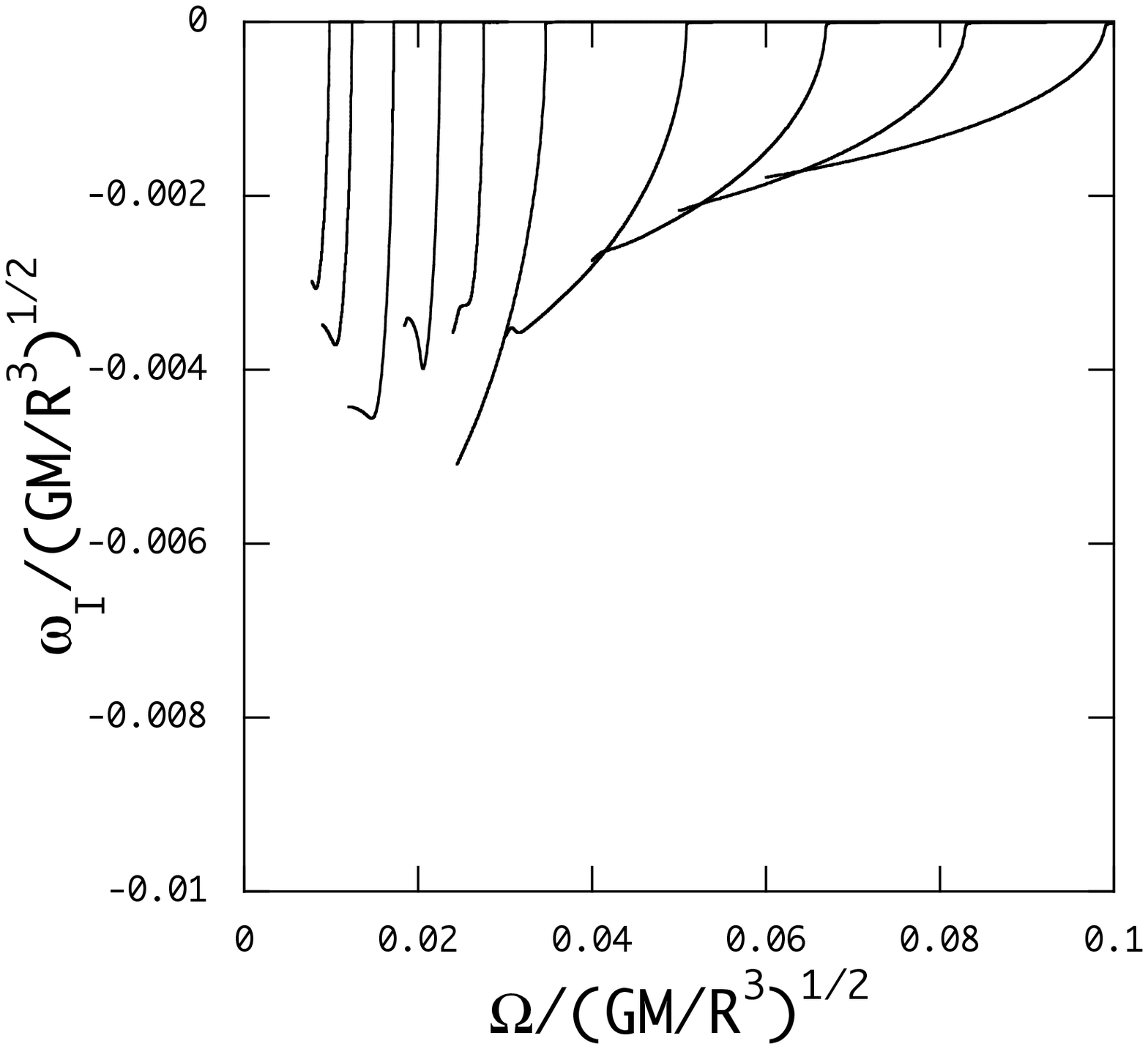}}
\resizebox{0.5\columnwidth}{!}{
\includegraphics{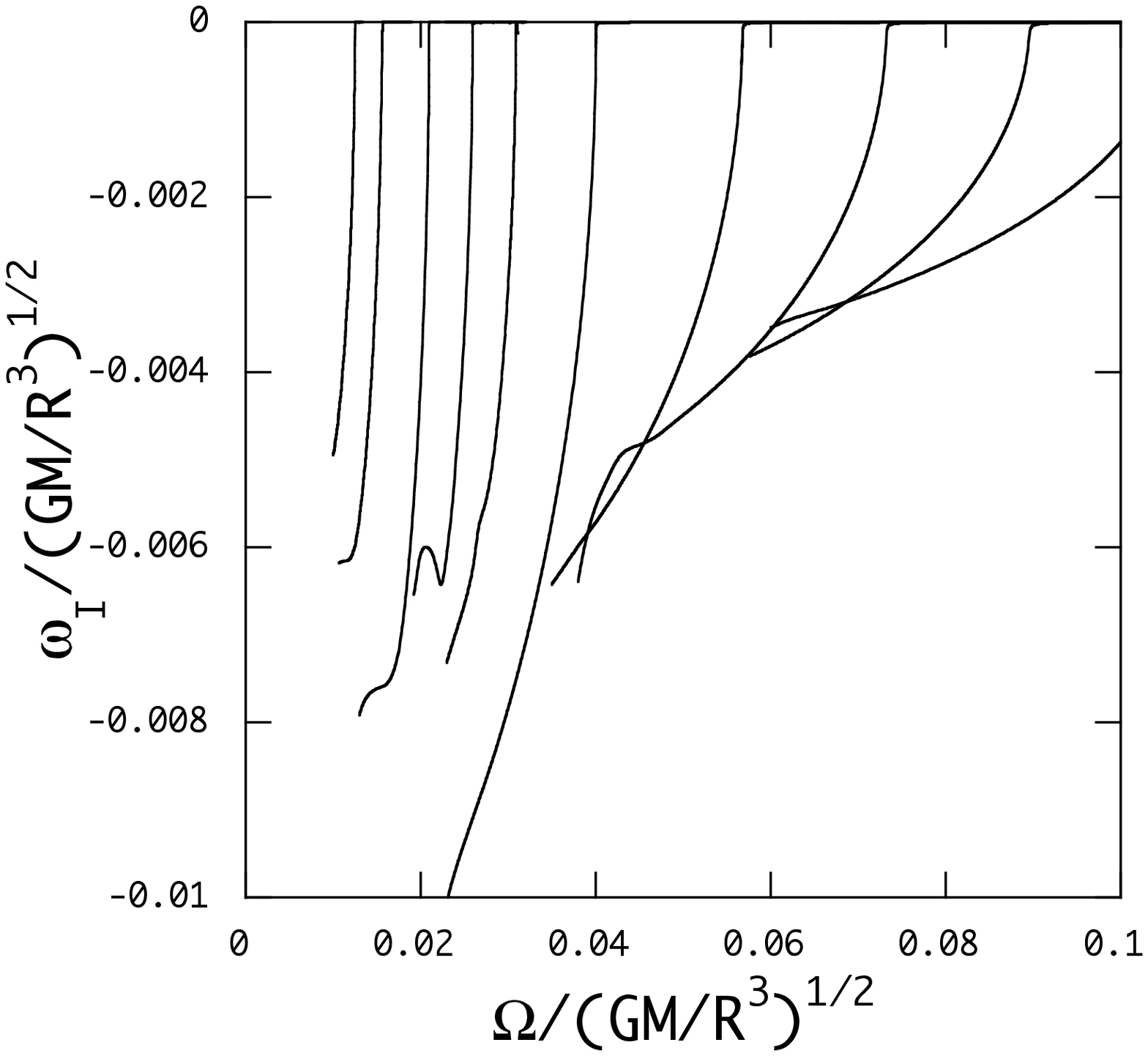}}
\label{fig:omegam4}
\caption{Complex frequency $\bar\omega=\omega/\sqrt{GM/R^3}$ of overstable convective modes of even parity for $m=-1$ and $m=-2$
as a function of $\bar\Omega=\Omega/\sqrt{GM/R^3}$ where $\delta_A=10^{-4}$ is assumed for the convective core.}
\end{figure}

The Jovian models computed in \S 2.1 have the structure of a polytrope of the index $n=1$ in the core
and of an isothermal atmosphere in the envelope.
For $\Gamma_1=2$, the quantity $|\Gamma_1^{-1}-d\ln\rho/d\ln p|$ in the core
is very small except in the layers close to the boundary between the radiative envelop and convective core.
Since $\Gamma_1^{-1}-d\ln\rho/d\ln p=\alpha_T\left(\nabla-\nabla_{\rm ad}\right)$,
$\Gamma_1^{-1}-d\ln\rho/d\ln p\approx 0$ means that the convective core
has an almost adiabatic structure.
If the planets are rapidly rotating and have magnetic fields, however,
finite negative values of $N^2$ in the convective core are more plausible (e.g., Stevenson 1979).
In this paper, assuming rapidly rotating Jovian planets, we expect $\Gamma_1^{-1}-d\ln\rho/d\ln p$ to have a finite positive value in the convective core, that is,
\be
{1\over\Gamma_1}-{d\ln\rho\over d\ln p}=\delta_A,
\ee
where $\delta_A$ is a positive constant.
Since $\alpha_T=1$ for an ideal gas, $\delta_A$ is equal to the superadiabatic temperature gradient $\nabla-\nabla_{ad}$.
For a positive value for $\delta_A$, we obtain
\be
N^2=-{g\over r}V\delta_A<0
\ee
in the convective core.
In the isothermal radiative envelope, we simply assume $\nabla_{ad}=0.5$ and $\nabla=0$.

In Figure 2, we plot several sequences of the complex eigenfrequency $\bar\omega=\bar\omega_{\rm R}+\rmi\bar\omega_{\rm I}$ for the $m=-1$ and $m=-2$ convective modes of even parity, assuming
$\delta_A=10^{-4}$ in the convective core, where
we have used $j_{\rm max}=12$ for the expansion of the perturbations.
Here, we are interested in prograde ($\bar\omega_{\rm R}>0$) and unstable ($\bar\omega_{\rm I}<0$)
convective modes of rotating Jovian planets (see Lee \& Saio 1986).
Note that convective modes at $\bar\Omega=0$ have pure imaginary eigenvalue $\bar\omega$ such that $\bar\omega^2<0$,
which is proportional to $l(l+1)$ with $l$ being the harmonic degree of the modes.
Each of the mode sequences, as a function of $\bar\Omega$, has a sharp peak of $\bar\omega_{\rm R}$, at which
the convective mode is stabilized in the sense that $|\bar\omega_{\rm I}|\ll\bar\omega_{\rm R}$.
The height of the peaks for $m=-2$ is higher that that for $m=-1$.
When the convective modes reach the peaks, they are resonantly coupled with envelope $g$-modes and
$\bar\omega_{\rm R}$ of the modes
starts decreasing rapidly with increasing $\bar\Omega$, keeping $\bar\omega_{\rm I}<0$, that is,
the convective modes stay unstable even in the 
presence of strong dissipations associated with $g$-modes in the envelope.
We have carried out similar calculations assuming $\delta_A=10^{-2}$ to find that the frequency $\bar\omega$ of
the convective modes scales as $\bar\omega\propto\sqrt{\delta_A}$.

We find that convective modes in the rapidly rotating planets are separated into two types, which we may call type A and type B.
For the convective modes belonging to type A, the complex frequency $\bar\omega$
is approximately given by equation (\ref{eq:wkbomegar}), which is derived based on the WKB analysis discussed in the Appendix. 
We can therefore assign a pair of indices $(k,n)$ to each of the mode sequences, where
the index $k$ is an integer used to label the eigenvalue $\lambda_{k,m}$ of the Laplace tidal
equation and the index $n$ is also an integer corresponding to the number of radial nodes of the dominant
expansion coefficient $S_l(r)$ of the eigenfunction (see Appendix).
For the stabilized convective modes, the integer index $k$ is negative and $\lambda_{k,m}$ becomes negative
for prograde modes of $\nu\equiv 2\Omega/\omega_{\rm R}>0$ (see Fig. \ref{fig:lambda}).
The examples of such assignment of the indices $(k,n)$ to type A convective modes are given in Fig. 3.
Note that we found only convective modes that could be associated with $k=-2$, $-6$, $-10$, $\cdots$.
For the convective modes of type A, the angular part of the eigenfunction is approximately represented by
the Hugh function $\Theta_{k,m}(\cos\theta)$, which is an eigenfunction of the Laplace tidal equation associated with the eigenvalue $\lambda_{k,m}$
and can be represented by a linear combination of the associated Legendre functions $P_l^m(\cos\theta)$ (e.g., Lee \& Saio 1997).
When $|\bar\omega_{\rm I}|\ll \bar\omega_{\rm R}$, the stabilized convective modes are likely to
interact with gravity waves propagating in the envelope.
The interactions affect the complex frequency $\bar\omega$ and may produce wiggles in $\omega_{\rm I}$ as a function of $\bar\Omega$.

\begin{figure}
\resizebox{0.45\columnwidth}{!}{
\includegraphics{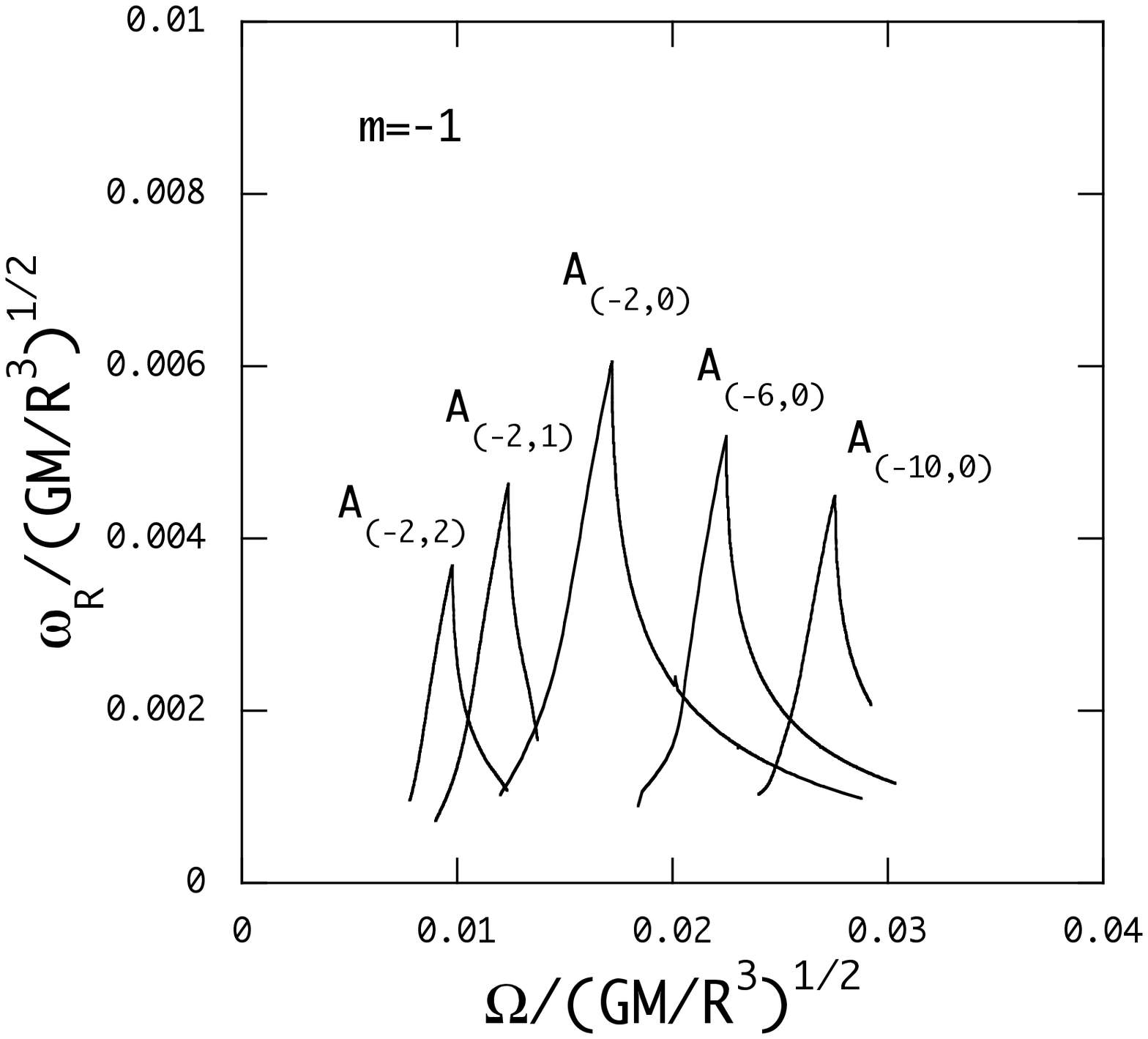}}
\resizebox{0.45\columnwidth}{!}{
\includegraphics{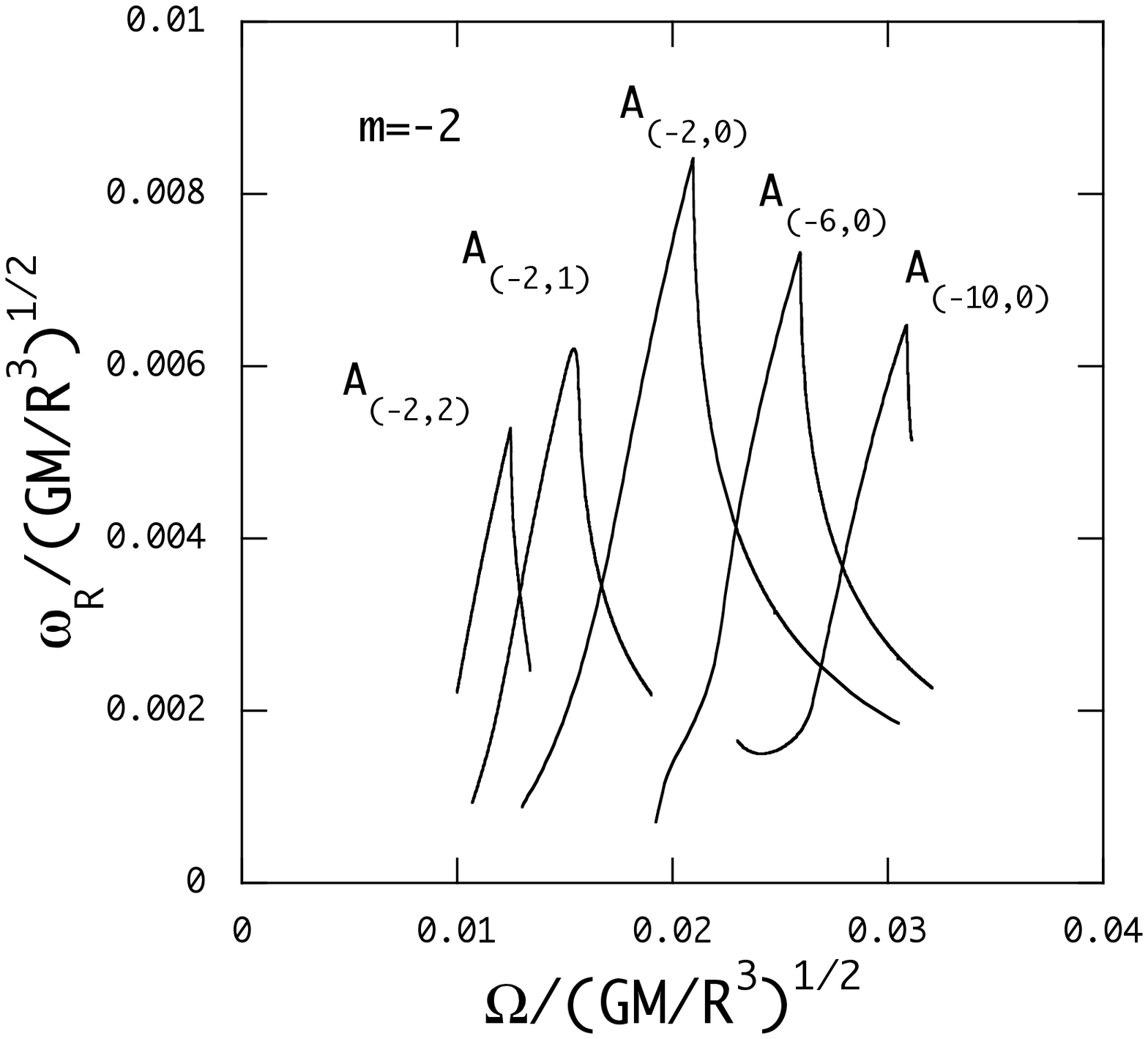}}
\label{fig:omegar_km}
\caption{Examples of the assignment of $(k,n)$ indices to the type A overstable convective modes
in Fig. 2.}
\end{figure}

Let $\bar\Omega_{k,m,n}$ signify the rotation frequency at which type A convective modes
are stabilized making a peak of $\bar\omega_{\rm R}$.
From Fig. 3, we find $\bar\Omega_{-2,-2,0}>\bar\Omega_{-2,-1,0}$.
We also find that $\bar\Omega_{-2,m,n}<\bar\Omega_{-6,m,n}<\bar\Omega_{-10,m,n}<\cdots$ for given $m$ and $n$, and
the number of nodes in the $\theta$ direction increases with increasing $|k|$ for negative $k$.
This indicates that as $|k|$ and $|m|$ increase, the horizontal wavelengths become shorter and
the rotation rate necessary to stabilize convective modes becomes higher.
This trend is opposite to that for the wavelengths in the radial direction, that is,
for given $k$ and $m$, $\bar\Omega_{k,m,n}$ decreases as the wavelengths in the radial direction decrease
with increasing $n$.

For the convective modes of type B, on the other hand, 
the first component $S_l$ of $l=|m|$ has dominant amplitudes, particularly
before their stabilization.
To classify the type B convective modes, we simply use the number of radial nodes of the eigenfunction such that
$B_n$.
To stabilize the type B convective modes, we generally need much higher rotation rates $\bar\Omega$ than
for the type A convective modes.
The rotation rate $\bar\Omega$
necessary to stabilize the type B convective modes increases as the number of radial nodes increases, which trend is
opposite to that found for the type A convective modes.

Because of strong non-adiabatic effects in the envelope, 
the kinetic energy of unstable convective modes dissipates and can be used to heat up the thin envelope.
The normalized energy dissipation rate, $D$, averaged over the oscillation period, may be given by
\be
D&\equiv&-{1\over L_{eq}P}\int_0^P{dE\over dt} dt\nonumber\\
&\approx&{\bar\omega_{\rm R}\over 2}\int_0^{R}{\rm Im}\left(\sum_l{\delta T^*_l\over T}{\delta s_l\over c_p}\right)
{\rho Tc_pr^3\sigma_0\over L_{\rm eq}}{dr\over r}\nonumber\\
&=&-{\bar\omega_{\rm R}\over 2}\int_0^{R}{\rm Im}\left[{1\over\rmi\omega/\omega_D+1}\sum_l\nabla_{ad}{\delta p_l^*\over p}\left(\nabla_{ad}{\delta p_l\over p}+\nabla VS_l\right)\right]
{\rho Tc_pr^3\sigma_0\over L_{\rm eq}}{dr\over r},
\label{eq:defd}
\ee
where $P=2\pi/\omega_{\rm R}$, $L_{\rm eq}=4\pi R^2 F_{\rm eq}$, $x=r/R$,
$\delta T/T=\nabla_{ad}\delta p/p+\delta s/c_p$, and we have assumed $|\omega_{\rm I}|\ll|\omega_{\rm R}|$. 
Assuming that the host star has the effective temperature $T_{\rm eff,*}=5.8\times10^3$K, the radius $R_*=R_\odot$,
and the distance $d$ between the planet and the host star is $d=0.05$AU, we estimate $L_{eq}=4\pi R^2\sigma_{\rm SB}T^4_{\rm eff,*}(R_*/d)^2\approx 6\times10^{29}{\rm erg/s}$
where $\sigma_{\rm SB}$ is the Stefan Boltzmann constant.
Note also that if the spin of the planet is in synchronization with the orbital motion around the host star
so that $\Omega \approx \sqrt{GM_*/d^3}$,
we have $\Omega/\sigma_0\approx 0.05$ for the mass $M_*\approx M_\odot$ of the host star.

Assuming that the oscillation energy of the convective modes is approximately constant and
independent of $\Omega$, that is,
\be
\omega_{\rm R}^2\int_0^R\rho\pmb{\xi}\cdot\pmb{\xi}^*dV=\sigma_0^2f_0^2\int_0^R\rho r^2 dV,
\label{eq:norm}
\ee
where $f_0$ is the amplitude parameter, we define the non-dimensional oscillation amplitudes $A_c$ as
\be
A_c=\sqrt{\left(\int_0^R\rho r^2dr\right)^{-1}\int_0^R\left|\pmb{\xi}/r\right|^2\rho r^2 dr}.
\ee
The normalization (\ref{eq:norm}) implies that $A_c= f_0/|\bar\omega_{\rm R}|$ and hence
$f_0/|\bar\omega_{\rm R}|\ll 1$ for linear pulsations.
Replacing $R$ by $r$ in equation (\ref{eq:defd}), we may define the function $D(r)$,
which is plotted in Fig. 4 for the stabilized convective modes of $m=-1$ and $m=-2$ where we assume $f_0=1$.
Note that $D(r)$ and $D$ are proportional to $f_0^2$.
This figure shows that the strong heating takes place at the bottom of the radiative envelope.
In Figure 5, we plot the quantity $D$ as a function of $\bar\Omega$ along
the convective mode sequences of $m=-1$ and $m=-2$ for $\delta_A=10^{-4}$ where $f_0=1$ is used.
The magnitudes of $D$ for the type A convective modes are at most of order of 10 and
$D$ changes erratically as a function of $\bar\Omega$ after their stabilization.
These erratic behaviors of $D$ after stabilization are caused by interaction between
the convective modes in the core and gravity waves in the envelope.
For the type B convective modes, on the other hand, the magnitudes of $D$ after their stabilization
can be as large as $D\gtrsim 10^2$.
This rapid increase of $D$ indicates that energy dissipations associated with envelope $g$-modes 
become significant when $g$-modes are resonantly excited by convective modes after their stabilization.
Note that $D$ of the type B convective modes does not strongly suffer erratic changes even after
their stabilization.

\begin{figure}
\resizebox{0.45\columnwidth}{!}{
\includegraphics{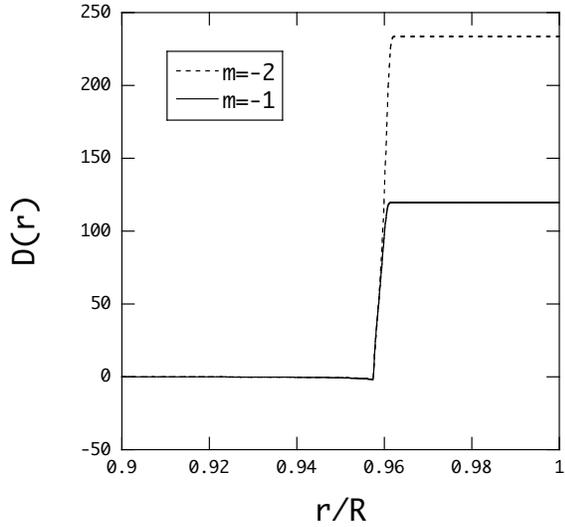}}
\label{fig:dr_m1m2}
\caption{$D(r)$ versus $x=r/R$ for the convective modes of $m=-1$ (solid line) and $m=-2$ (dotted line)
for $f_0=1$,
 where
$\bar\omega=(1.66\times10^{-3},-5.16\times10^{-7})$ for $m=-1$ and
$\bar\omega=(3.53\times10^{-3},-1.38\times10^{-6})$ for $m=-2$ at $\bar\Omega=0.05$.}
\end{figure}

\begin{figure}
\resizebox{0.45\columnwidth}{!}{
\includegraphics{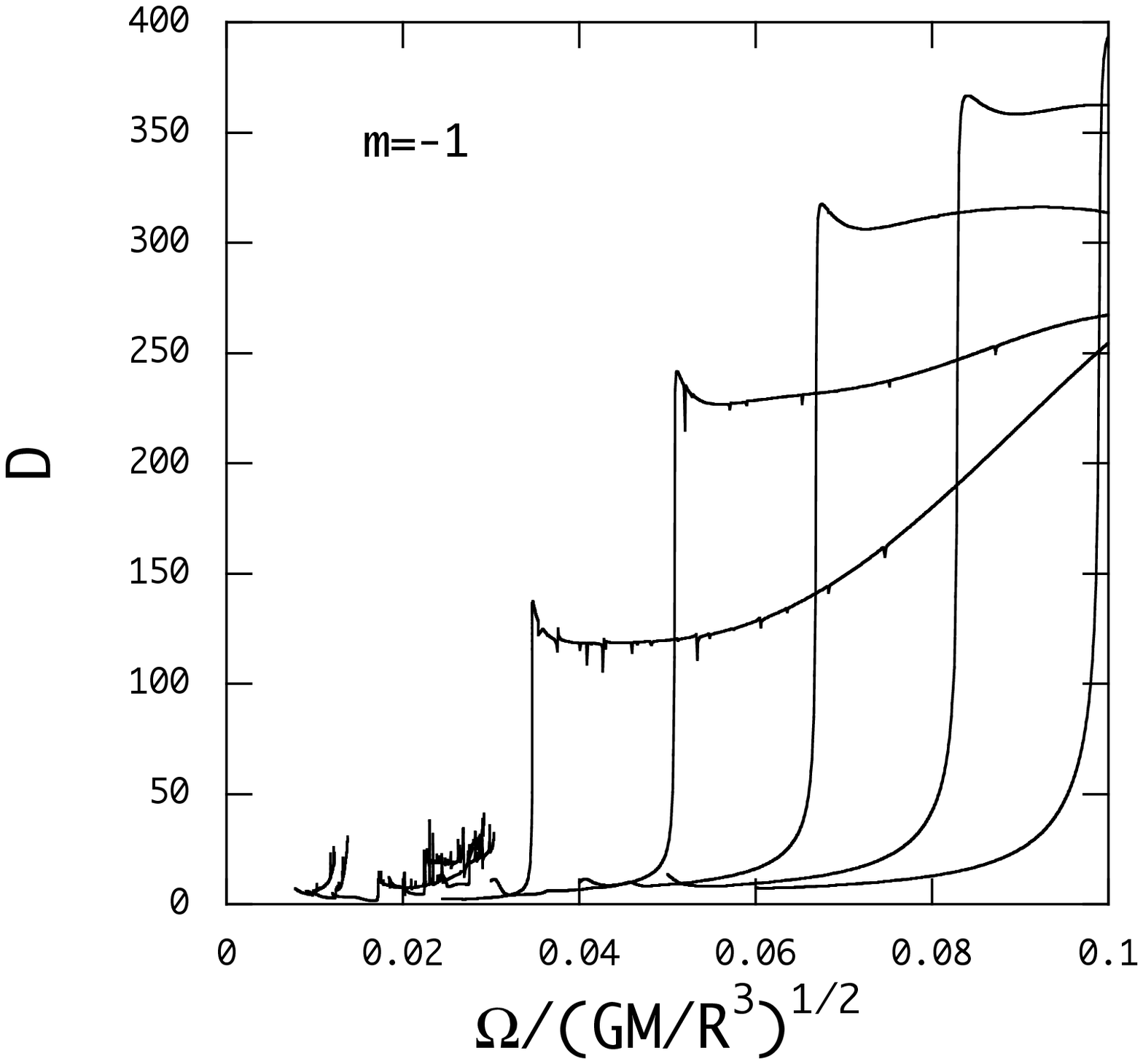}}
\resizebox{0.45\columnwidth}{!}{
\includegraphics{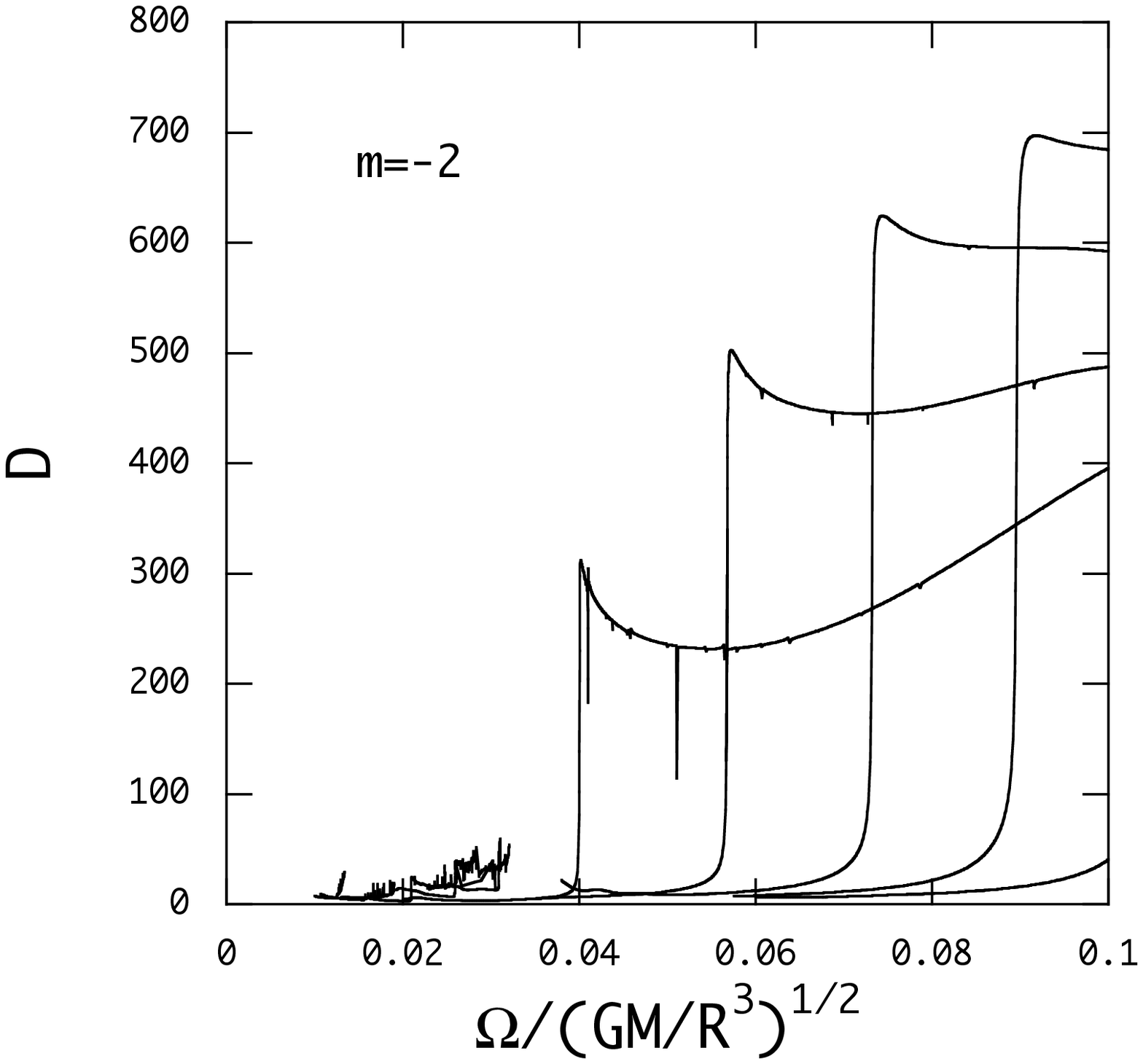}}
\label{fig:ddx}
\caption{Dissipation rate $D$ along the overstable convective mode sequences of $m=-1$ (left panel) and $m=-2$ (right panel),
where $f_0=1$ is assumed for the amplitude normalization (\ref{eq:norm}).}
\end{figure}

\section{Conclusions}

We calculate overstable convective modes in rotating hot Jupiters that
have the convective core and thin radiative envelope, by
taking account of the non-adiabatic effects.
As the rotation rate increases, the convective modes in the core are stabilized to attain $|\bar\omega_{\rm I}|\ll\bar\omega_{\rm I}^0$, where $\bar\omega^0_{\rm I}$ is the value of $\bar\omega$ at $\bar\Omega=0$.
For the stabilized convective modes we also have $|\bar\omega_{\rm I}|\ll|\bar\omega_{\rm R}|$.
We find that the stabilized convective modes are separated into two types, A and B, depending on the angular dependence of the eigenfunctions.
The qualitative behavior of type A convective modes are well understood by the WKB analysis
under the traditional approximation for the oscillation modes of rotating stars.
The stabilized convective modes in the core excite gravity waves in the envelope and remain unstable even in the presence of strong non-adiabatic dissipations in the envelope.
This mechanism can be an excitation mechanism of low frequency prograde $g$ modes of
rotating stars having a convective core and a radiative envelope, as originally suggested for 
massive main sequence stars (Lee \& Saio 1986; Lee 1988).

Computing the non-adiabatic dissipation rate $D$ of the oscillation energy of the convective modes,
we showed that the dissipative heating most strongly occurs at the bottom of the envelope and that
$D$ becomes comparable to or even greater than the irradiated luminosity due to the host star
for the oscillation amplitudes $A_c\sim f_0/\bar\omega_{\rm R}$ for $f_0=1$.
Baraffe et al (2003) carried out evolution calculation of irradiated Jovian planets
to obtain inflated hot Jupiter models.
They showed that if there exists an extra heat source of the magnitudes of order of $L_{extra}\sim10^{27}-5\times10^{27}~{\rm erg/s}$ in the deep interior, the planet models could inflate
by an enough amount to attain the radii consistent with the observationally estimated radius of the hot Jupiter HD 209458b.
For $L_{eq}\sim 5\times 10^{29}~{\rm erg/s}$ and hence $L_{extra}/L_{eq}\sim 10^{-2}$,
to produce the heating rate $L_{extra}\sim D_1f_0^2L_{eq}$ where $D_1$ is the value of $D$ for $f_0=1$ and
$D_1\sim 1-10^3$ as suggested by Fig. 5,
we can use $f_0\sim 10^{-2}-10^{-1}$, for which
$A_c\gtsim 1$ for the convective mode frequency $\bar\omega_{\rm R}\sim 10^{-3}$.
In this case the linearity condition $A_c\ll 1$ could not be satisfied.
This suggests that the heating by the overstable convective modes could not be large enough to inflate the hot Jupiters so long as the amplitudes are limited to the linear regime.

The conclusion given above concerning the magnitudes of $D$ has been derived based on a very approximate treatment of non-adiabatic effects on the oscillation modes.
For the non-adiabatic analyses of oscillation modes of hot Jupiters, it is desirable to use
evolutionary models computed with appropriate opacities and equations of state.
Only with such Jovian models, we are able to reach a definite conclusion concerning the efficiency of the heating mechanism we discussed in this paper.

The frequency $\bar\omega$ of convective modes and the rotation frequency $\bar\Omega_{k,m,n}$
at which the convective modes are stabilized are approximately proportional to 
$\sqrt{\delta_A}$ assumed for the convective core.
The value of $\delta_A=10^{-4}$, which we use in this paper, 
may be large for the interior of rotating Jovian planets as suggested by Stevense (1976).
Even for a much smaller value of $\delta_A$, convective modes with large $|k|$ and $|m|$ (type A)
or with large $n$ and $|m|$ (type B) could survive rotational stabilization.
The presence of a magnetic field may change the conclusion about the rotational stabilization of convective modes.
It is an important problem to numerically examine the effects of rotation and 
magnetic field on the stabilization of convective modes of rotating stars.

\appendix
\section{WKB analysis in the traditional approximation}

\begin{figure}
\resizebox{0.45\columnwidth}{!}{
\includegraphics{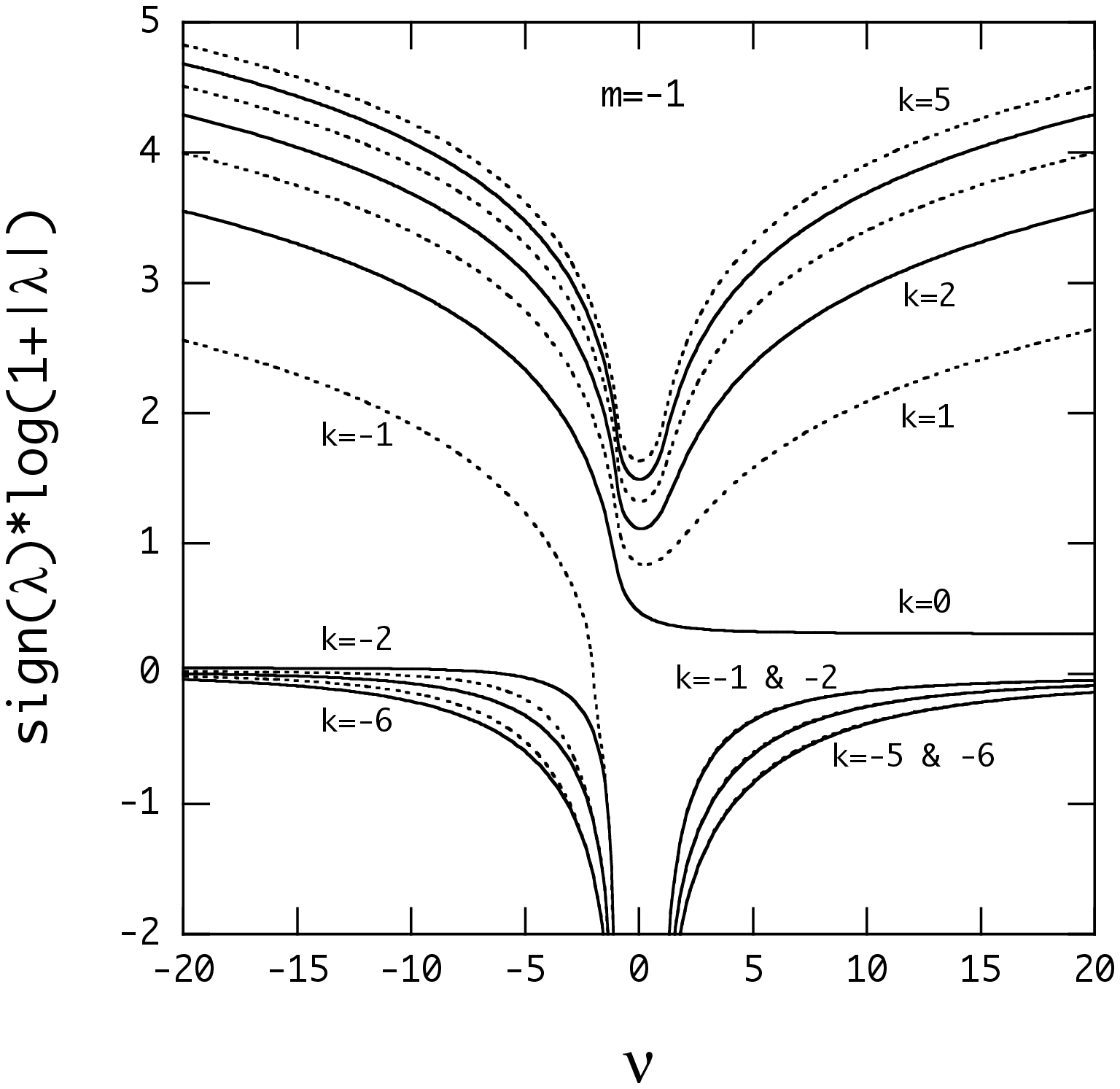}}
\resizebox{0.473\columnwidth}{!}{
\includegraphics{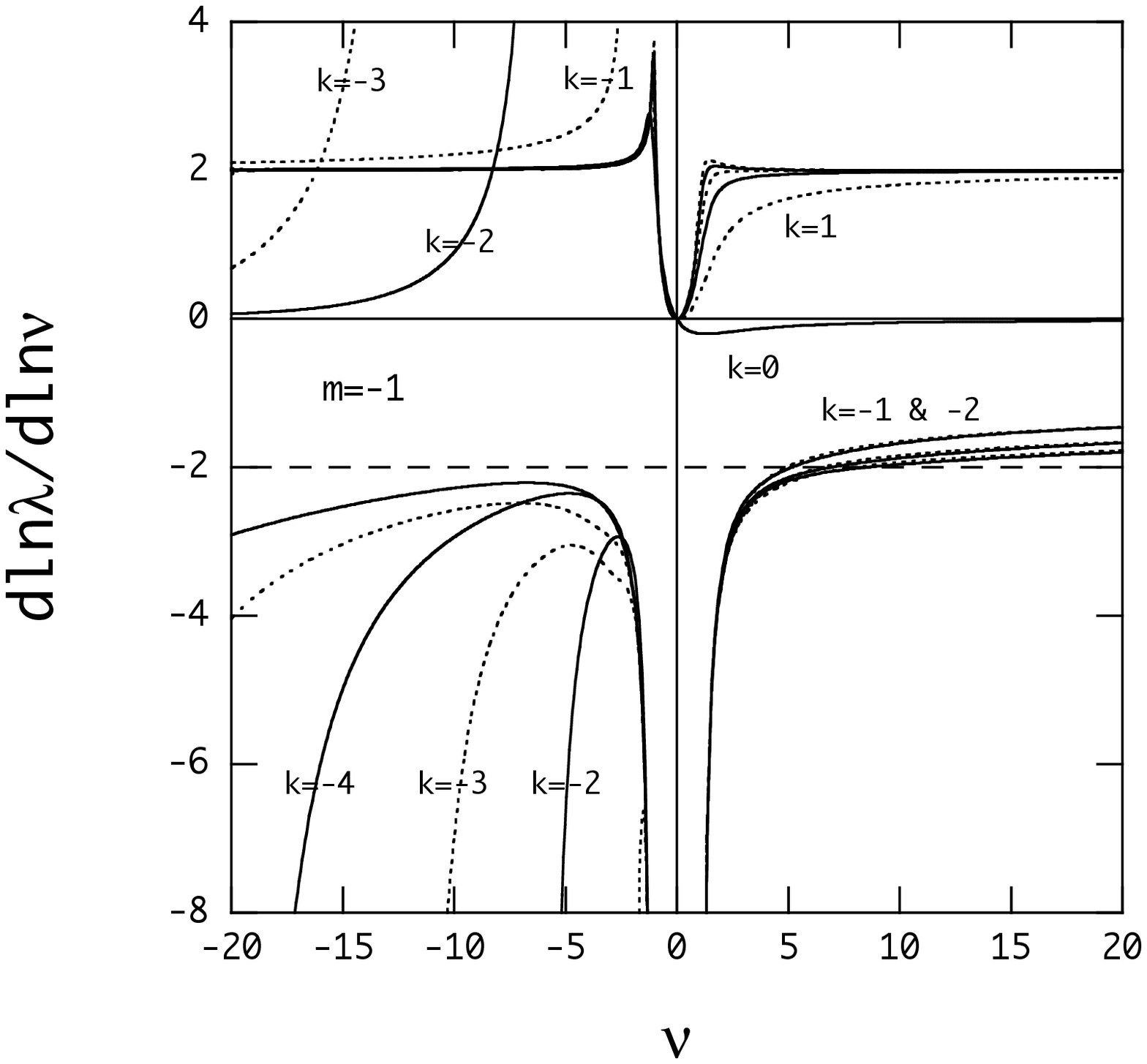}}
\caption{Eigenvalue $\lambda_{k,m}$ (left panel) and its derivative $d\ln\lambda_{k,m}/d\ln\nu$ (right panel) for $m=-1$
where $\nu=2\Omega/\omega$.}
\label{fig:lambda}
\end{figure}

WKB analysis is a useful tool to understand the numerical results presented in \S 3.
We employ the traditional approximation to discuss the oscillations of uniformly rotating stars,
neglecting the terms proportional to $\Omega_H=-\Omega\sin\theta$ in the perturbed equations of motion.
In the traditional approximation, the $\theta$ dependence of the 
perturbations such as $\xi_r$ and $p'$ can be represented by a single function $\Theta_{k,m}(\theta)$ so that
$\xi_r(r,\theta,\phi,t)=\xi_r(r)\Theta_{k,m}(\theta){\rm e}^{\rmi(m\phi+\sigma t)}$, where
$\Theta_{k,m}(\theta)$ is called the Hough function, which is an eigenfunction of the Laplace's tidal equation associated with the eigenvalue $\lambda_{k,m}$ and $k$ is an integer introduced to classify the oscillation modes
 (see, e.g., Lee \& Saio 1997).
The eigenvalue $\lambda_{k,m}$ depends on $\nu=2\Omega/\omega$ for given $(k,m)$.
In Figure \ref{fig:lambda}, $\lambda_{k,m}$ and its derivative $d\ln\lambda_{k,m}/d\ln\nu$
are plotted as a function of $\nu=2\Omega/\omega$ for $m=-1$, where positive (negative) $\nu$
corresponds to prograde (retrograde) modes for negative $m$.
For non-negative integers $k$, 
$\lambda_{k,m}\rightarrow l_k(l_k+1)$ with $l_k=|m|+k$ when $\nu\rightarrow 0$, and 
$\lambda_{k,m}\propto \nu^2$ for large $|\nu|$ except for $\lambda_{k=0,m}$ on the prograde side ($\nu>0$).
For negative integers $k$, $\lambda_{k,m}$ on the retrograde side vanishes at $\nu=l'_k(l'_k+1)/m$ with $l'_k=|m|+|k|-1$ and positive $\lambda_{k,m}$s on the retrograde side are associated with 
$r$-modes.
Similar figures for $m=-2$ are given in Lee \& Saio (1997).

As discussed in Lee \& Saio (1989), the coupling between rotationally stabilized convective modes
of $k_c<0$ in the core and low frequency $g$-modes of $k_g\ge0$ in the envelope of a massive main sequence star may be described by a dispersion relation
\be
\tan\Phi_c(\omega)\tan\Phi_g(\omega)=\epsilon,
\ee
where $\epsilon$ is the coupling coefficient between the two modes, and 
\be
\Phi_c(\omega)=\int_0^{x_c}{\sqrt{-\lambda_{k_c,m}rA/c_1}\over\bar\omega}{dx\over x}-\alpha_{k_c}\pi, \quad
\Phi_g(\omega)=\int_{x_c}^1{\sqrt{-\lambda_{k_g,m}rA/c_1}\over\bar\omega}{dx\over x}-\alpha_{k_g}\pi,
\ee
where $x=r/R$ and $x_c$ represents the boundary between the convective core and radiative envelope,
and $\alpha_k$ is a constant which may depend of $k$ and $m$.
Note that for $m<0$ and $\nu>0$, $\lambda_{k_c,m}$ for the convective modes is negative and 
$\lambda_{k_g,m}$ for the $g$-modes is positive as shown by Figure \ref{fig:lambda}.
Assuming that gravity waves in the envelope are totally dissipated, we replace
$\tan\Phi_g=(e^{\rmi\Phi_g}-e^{-\rmi\Phi_g})/(e^{\rmi\Phi_g}+e^{-\rmi\Phi_g})\rmi$ for $g$-modes by $-\rmi$ to obtain
\be
\tan\Phi_c(\omega)=\epsilon \rmi.
\label{eq:disp2}
\ee
For complex frequency $\omega=\omega_{\rm R}+\rmi\omega_{\rm I}$, assuming $|\omega_{\rm I}|\ll|\omega_{\rm R}|$, we obtain
\be
\tan\Phi_c(\omega_{\rm R})+
\rmi\omega_{\rm I}{\partial\over\partial\omega_{\rm R}}\tan\Phi(\omega_{\rm R})
=\epsilon\rmi,
\label{eq:dispersionrelation}
\ee
the real part of which gives
\be
\tan\Phi_c(\omega_{\rm R})=0,
\ee
and determines the frequency as
\be
\bar\omega_{\rm R}={1\over(\alpha_{k_c}+n)\pi}\int_0^{x_c}\sqrt{-\lambda_{k_c,m}{rA\over c_1}}{dx\over x},
\label{eq:wkbomegar}
\ee
where $n$ is an integer that satisfies $\Phi_c(\omega_{\rm R})=n\pi$.
On the other hand, the imaginary part of equation (\ref{eq:dispersionrelation}) leads to
\be
\omega_{\rm I}{\partial\over\partial\omega_{\rm R}}\tan\Phi_c(\omega_{\rm R})
=-{\omega_{\rm I}\over\omega_{\rm R}}\left(\int_0^{x_c}{\sqrt{-\lambda_{k_c,m}{rA/ c_1}}\over\bar\omega_{\rm R}}{dx\over x}\right)F_{k_c,m}=\epsilon
\ee
and hence
\be
\omega_{\rm I}=-{\omega_{\rm R}\epsilon\over(\alpha_{k_c}+n)\pi F_{k_c,m}}
\ee
where
\be
F_{k,m}=1+{1\over 2}{d\ln|\lambda_{k,m}|\over d\ln\nu}.
\ee
We let $\nu^0$ satisfy $F_{k,m}(\nu^0)=0$, which occurs only for 
prograde modes with negative $k$.
Since
\be
F_{k,m}{\partial\bar\omega_{\rm R}\over\partial\bar\Omega}={d\ln|\lambda_{k,m}|\over d\nu},
\ee
we have ${\partial\bar\omega_{\rm R}/\partial\bar\Omega}\rightarrow\pm\infty$ when
$\nu\rightarrow \nu^0$.
Using WKB analysis, Lee \& Saio (1989, 1990, 1997) discussed that
the energy of the oscillation modes of rotating stars
is proportional to $\lambda_{k,m} F_{k,m}$ and the energy can be negative for rotationally stabilized convective modes associated with $\lambda_{k,m}<0$ if $F_{k,m}>0$.
As shown by Figure \ref{fig:lambda}, $F_{k,m}$ for the convective modes becomes positive
for $\nu\gtsim 5$.
The convective modes appear as prograde modes of $\omega_{\rm R}>0$ for 
$m<0$
and hence have $\omega_{\rm I}<0$ for $F_{k,m}>0$, that is, the convective modes with
negative energy become overstable when the oscillation energy leaks out of the core.

\begin{table*}
\begin{center}
\caption{$\nu^0$ and $-\lambda^0_{k,m}\nu^0_{k,m}$ for convective modes for $m=-1$ and $m=-2$.}
\begin{tabular}{@{}ccccc}
\hline
  &  $m=-1$ && $m=-2$\\
\hline
 $k$ & $\nu^0_{k,m}$ & $-\lambda^0_{k,m}\nu^0_{k,m}$ & $\nu^0_{k,m}$ &$-\lambda^0_{k,m}\nu^0_{k,m}$\\
\hline
$-1$ & 4.92 & 6.37 &4.66 & 13.3\\
$-2$ & 5.15 & 6.29 &4.70 & 13.3\\
$-3$ & 6.52 & 11.5 &5.79 & 20.6\\
$-4$ & 6.94 & 11.2 &5.88 & 20.5\\
$-5$ & 7.93 & 17.4 &6.83 & 28.6 \\
$-6$ & 8.52 & 16.7 &6.96 & 28.3\\
$-7$ & 9.23 & 23.9 &7.79 & 37.3\\
$-8$ & 9.96 & 22.8  &7.98 & 36.9\\
$-9$ & 10.4 & 30.9 &8.70 & 46.7\\
$-10$ & 11.3 & 29.3 & 8.94 & 45.9\\
\hline
\end{tabular}
\medskip
\end{center}
\end{table*}

At $\nu=\nu^0_{k,m}$ the real part of the frequency $\omega_{\rm R}^0$ is given by equation 
(\ref{eq:wkbomegar}) and hence the rotation rate $\Omega^0=\nu^0\omega_{\rm R}^0/2$.
We thus have for convective modes of negative $\lambda_{k,m}$ and $k$
\be
\bar\omega^0_{\rm R}\bar\Omega^0=\nu^0(\bar\omega^0_{\rm R})^2/2={1\over 2\pi^2}{-\lambda^0_{k,m}\nu^0_{k,m}\over
(\alpha_k+n)^2}\left(\int_0^{x_c}\sqrt{rA\over c_1}{dx\over x}\right)^2,
\ee
where $\lambda^0_{k,m}=\lambda_{k,m}(\nu^0_{k,m})$.
Assuming
\be
{1\over 2\pi^2}{-\lambda^0_{k,m}\nu^0_{k,m}\over
(\alpha_k+n)^2}=c^2
\ee
with $c$ being a constant, we obtain for $n=0$
\be
\alpha_k={1\over\sqrt{2}\pi c}\sqrt{-\lambda^0_{k,m}\nu^0_{k,m}},
\ee
and the value of $-\lambda^0_{k,m}\nu^0_{k,m}$ is tabulated in Table 1.

We numerically solve the dispersion relation (\ref{eq:disp2}) and pick up the convective modes
that satisfies $\bar\omega_{\rm R}>0$, $\bar\omega_{\rm I}<0$, $|\bar\omega_{\rm I}/\bar\omega_{\rm R}|<1/2$, and
$n\ge0$, where we use $\epsilon=10^{-3}$ and $\delta_A=10^{-4}$.
In Figure A2, we plot the complex frequency $\bar\omega$ for $m=-1$ convective modes of even parity.
We find a reasonable agreement between
the WKB results and the full numerical calculations shown in Figure 2 for type A convective modes, which probes
that the WKB analysis is useful to classify the convective modes 
numerically obtained.
Note however that the WKB equation (\ref{eq:wkbomegar}) cannot necessarily be used to explain the type B convective modes plotted in Fig. 2.

\begin{figure}
\resizebox{0.45\columnwidth}{!}{
\includegraphics{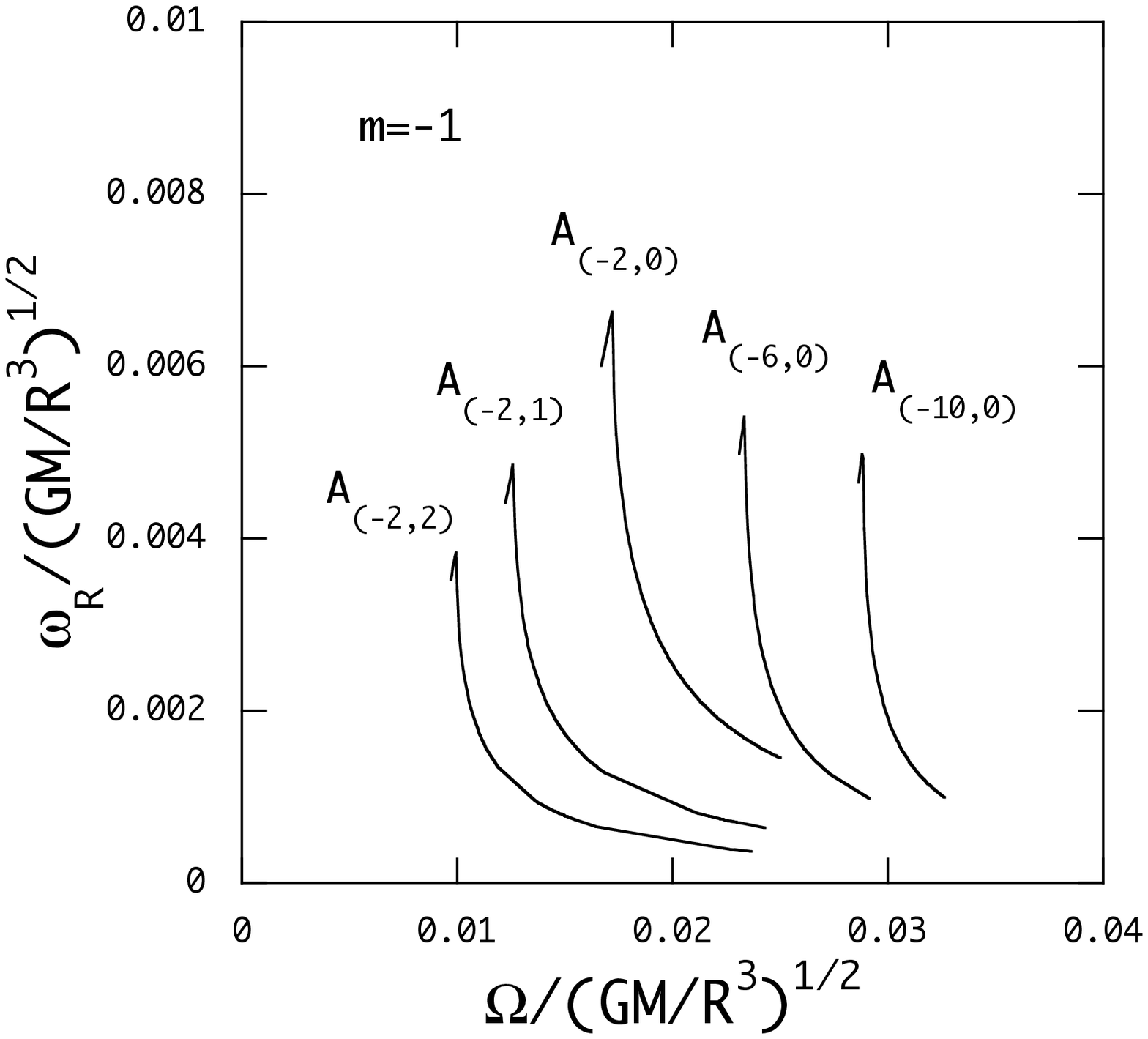}}
\resizebox{0.45\columnwidth}{!}{
\includegraphics{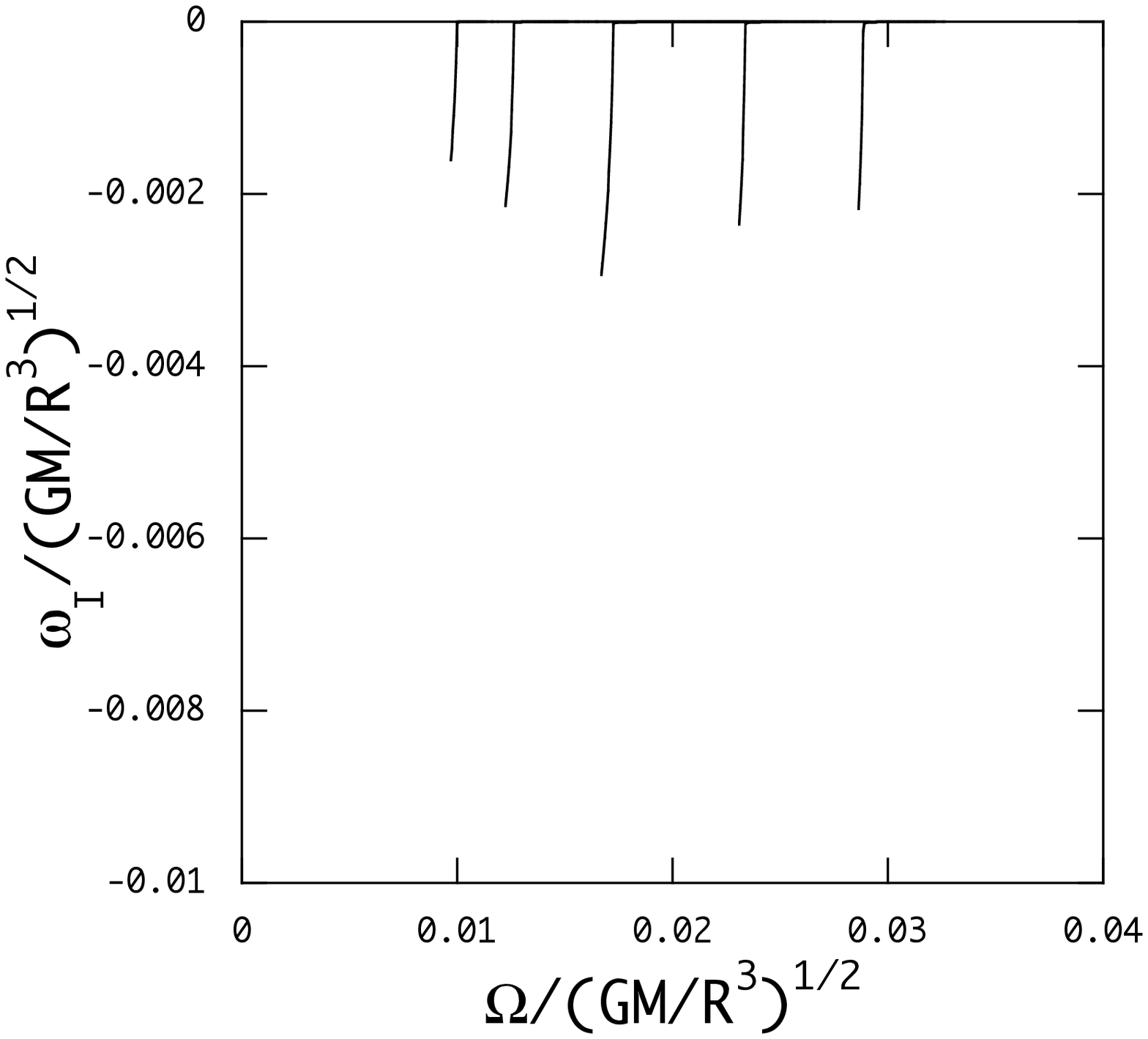}}
\label{fig:wkb_omega_mm1}
\caption{Complex frequency $\bar\omega$ of overstable convective modes as a function of $\bar\Omega$ for
$m=-1$ even modes for $\delta_A=10^{-4}$, where the modes are obtained by solving WKB dispersion relation (\ref{eq:disp2}).
The solid, dotted, and dashed lines represent the convective modes associated with $\lambda_{k,m}$ for
$k=-2$, $-6$, and $-10$, respectively, and the integer $n$ represents the radial order of the modes.}
\end{figure}


\begin{thebibliography}{}

\bibitem[Auclair-Desrotour P. et al(2017)]{2017A&A..603..A108} Auclair-Desrotour P. et al, 2017, A\&A, 603, A108
\bibitem[Auclair-Desrotour P., Leconte J.(2018)]{2018A&A..613..A45} Auclair-Desrotour P., Leconte J., 2018, A\&A, 613, A45
\bibitem[Arras P., Bildsten L.(2006)]{2006ApJ..650..394} Arras P., Bildsten L., 2006, ApJ, 650, 394
\bibitem[Arras P., Socrates A.(2010)]{2010ApJ..714..1} Arras P., Socrates A., 2010, ApJ, 714, 1
\bibitem[Baraffe I.etal(2003)]{2003A&A..402..701} Baraffe I., Chabrier G., Barman T.S., Allard F., Hauschildt P.H., 2003, A\&A, 402, 701
\bibitem[Deming D.etal(1989)]{1989ApJ..343..456} Deming D., Mumma M.J., Espenak F., Jennings D.E., Kostiuk T., Wiedemann G.,
Loewenstein R., Piscitelli J, 1989, ApJ, 343, 456
\bibitem[Iro N.etal(2005)]{2005A&A..436..719} Iro N., B\'ezard B., Guillot T., 2005, A\&A, 436, 719

\bibitem[Lee U., Saio H.(1986)]{1986MNRAS..221..365} Lee U., Saio H., 1986, MNRAS, 221, 365
\bibitem[Lee U., Saio H.(1987a)]{1987aMNRAS..224..513} Lee U., Saio H., 1987a, MNRAS, 224, 513
\bibitem[Lee U., Saio H.(1987b)]{1987bMNRAS..225..643} Lee U., Saio H., 1987b, MNRAS, 225, 643
\bibitem[Lee U., Saio H.(1989)]{1989MNRAS..237..875} Lee U., Saio H., 1989, MNRAS, 237, 875
\bibitem[Lee U., Saio H.(1990a)]{1990aApJ..359..29} Lee U., Saio H., 1990a, ApJ, 359, 29
\bibitem[Lee U., Saio H.(1990b)]{1990bApJ..360..590} Lee U., Saio H., 1990b, ApJ, 360, 590
\bibitem[Lee U., Saio H.(1997)]{1997ApJ..491..839} Lee U., Saio H., 1997, ApJ, 491, 839

\bibitem[Magalh\~aes J.A.etal(1989)]{1989Nature..337..444} Magalh\~aes J.A., Weir A.L., Conrath B.J., Gierasch P.J., Leroy S.S., 1989, Nature, 337, 444

\bibitem[Stevenson D.J.(1979)]{Geophys.Astrophy.Fluid.Dynamics1979..12..139} Stevenson D.J., Geophys. Astrophy. Fluid. Dynamics, 1979, 12, 139
\bibitem[Stevenson D.J.(1979a)]{1977ApJS..35..221} Stevenson D.J., Salpeter E.E., 1977, ApJS, 35, 221
\bibitem[Stevenson D.J.(1979b)]{1977ApJS..35..239} Stevenson D.J., Salpeter E.E., 1977, ApJS, 35, 239

\end{thebibliography}
\end{document}